\documentclass[%
preprint,
 amsmath,amssymb,
 aps,
 prl,
 longbibliography,
 lengthcheck,%
]{revtex4-1}

\usepackage{graphicx}
\usepackage{dcolumn}
\usepackage{bm}
\usepackage{hyperref}
\usepackage{mathtools}
\usepackage[mathlines]{lineno}

\def\ba{\begin{equation}\begin{array}{c}}
\def\ea{\end{array}\end{equation}}
\def\be{\ba\displaystyle}
\def\ee{\ea}

\newcommand{\ra}{\rangle}
\newcommand{\la}{\langle}

\renewcommand{\H}{\hat H}
\newcommand{\V}{\hat V}

\newcommand{\tr}{{\rm tr}}

\newcommand{\p}{\pmb p}

\newcommand{\ssigma}{\pmb \sigma}

\newcommand{\mfp}{\lambda_*}

\DeclareMathAlphabet{\mathcalligra}{T1}{calligra}{m}{n}

\begin{document}


\title{Timescale for adiabaticity breakdown in driven many-body systems \\ and
orthogonality catastrophe}

\author{Oleg Lychkovskiy$^{1,2,3}$}
\author{Oleksandr Gamayun$^{4,5}$}
\author{Vadim Cheianov$^{4}$}

\affiliation{$^1$ Skolkovo Institute of Science and Technology,
Skolkovo Innovation Center 3, Moscow  143026, Russia}
\affiliation{$^2$ Steklov Mathematical Institute of Russian Academy of Sciences,
Gubkina str. 8, Moscow 119991, Russia}
\affiliation{$^3$ Russian Quantum Center, Novaya St. 100A, Skolkovo, Moscow
Region, 143025, Russia}
\affiliation{$^4$ Instituut-Lorentz, Universiteit Leiden
P.O. Box 9506, 2300 RA Leiden, The Netherlands}
\affiliation{$^5$ Bogolyubov Institute for Theoretical Physics, 14-b
Metrolohichna str., Kyiv 03680, Ukraine}

\date{\today}


\begin{abstract}
The adiabatic theorem is a fundamental result in
quantum mechanics, which states that a system can be kept
arbitrarily close to the instantaneous ground state of its Hamiltonian if
the latter varies in time slowly enough.
The theorem has an impressive record of applications ranging from foundations
of quantum field theory to computational molecular dynamics. In light
of this success it is remarkable that a practicable quantitative understanding
of what {\it slowly enough} means is limited to a
modest set of systems mostly having a small Hilbert space.
Here we show how this gap can be bridged for a broad natural class of physical
systems, namely many-body systems where a small move in the parameter space induces an
orthogonality catastrophe. In this class,  the conditions for adiabaticity are
derived from the scaling properties of the parameter-dependent ground state
without a reference to the excitation spectrum.
This finding constitutes a major simplification of a complex problem, which
otherwise requires solving non-autonomous time evolution in a large Hilbert
space.

\end{abstract}

\maketitle


The adiabatic theorem (AT) is a profound statement that applies
universally to all quantum systems having slowly varying parameters.
It was originally conjectured by M.~Born in 1926 \cite{Born1926}, and
its complete proof was given in a joint paper by
by V.~Fock and M.~Born two years later \cite{born1928beweis}.
A number of refinements have been proposed over the years,
see  \cite{avron1999adiabatic} and references therein.
The theorem addresses the time evolution of
a generic quantum system having a Hamiltonian $\hat H_{\lambda},$
which is a continuous function of
a dimensionless time-dependent parameter
$\lambda = \Gamma t,$ where $t$ is time and $\Gamma$ is called the driving rate.
For each $\lambda$ one defines an instantaneous ground state,
which is the lowest eigenvalue solution to Schr\"odinger's stationary equation
\begin{equation}\label{eigenproblem}
\H_\lambda\, \Phi_\lambda = E_\lambda \, \Phi_\lambda.
\end{equation}
For simplicity, we assume
that $\Phi_\lambda$ is unique for each $\lambda.$
Imagine that at $t=0$ the system is prepared in the Hamiltonian's instantaneous
ground state
$\Phi_0.$ Then, as the parameter $\lambda$ changes with time,
the wave function  of the system, $\Psi_\lambda,$ evolves according to
Schr\"odinger's equation
\begin{equation}\label{Schrodinger equation}
i\,\Gamma\, \frac{\partial}{\partial \lambda} \Psi_\lambda = \H_{\lambda}\,
\Psi_\lambda, \qquad \Psi_0= \Phi_0.
\end{equation}
It is natural to expect that as time goes by, the physical state $\Psi_\lambda$
will depart from the instantaneous ground state $\Phi_\lambda,$ in other words
the quantum fidelity
\begin{equation}
 \mathcal F(\lambda) = \left | \langle \Phi_\lambda
 \vert \Psi_\lambda \rangle\right|^2,
 \label{Fdef}
\end{equation}
will decrease from its initial value of unity. The adiabatic
theorem states that this departure can be made arbitrarily small
provided that the driving is
slow enough. In more rigorous terms, for any $\lambda$
and for any small positive $\epsilon$ there exists small
enough $\Gamma$ such that $1-\mathcal F(\lambda)<\epsilon$.
A process in which the fidelity \eqref{Fdef} remains within a prescribed
vicinity of unity is called adiabatic.

The AT is a powerful tool in quantum physics, with applications ranging
from the foundations of perturbative quantum field theory
\cite{FetterWalecka,nenciu1989adiabatic}
to computational recipes in atomic and solid state physics
\cite{car1985unified}. Recent upsurge of interest in the AT
has been driven by the ongoing developments in the theory
of quantum topological order \cite{budich2013adiabatic} and quantum
information processing \cite{montanaro2016quantum}. The universal
applicability of the AT, however, comes at a cost.
Making no use of any specific properties of the Hamiltonian,
the AT's mathematical machinery does not provide a useful
definition of what is meant by ``slow enough.''
In particular, it leaves open the following
two questions (i) For a given displacement $\lambda$
in the parameter space what is the maximum driving
rate $\Gamma$ allowing to keep the evolution adiabatic?
and (ii) For a given driving rate $\Gamma$ what is the
system's {\it adiabatic mean free path,} that is
the maximum distance, $\lambda_*,$ in the parameter space that the system
can travel whilst maintaining adiabaticity?
With the advent
of technologies that depend on coherent quantum state manipulation
these questions are becoming of ever increasing practical importance.

For small or particularly simple systems questions (i) and (ii)
can be addressed microscopically, that is through
the explicit solution of Schrodinger's time-dependent equation \cite{Nakamura}.
The drawback of such a microscopic approach is that in larger
systems it stumbles upon the issues of computational complexity,
i.e. impossibility to solve the evolution in a huge Hilbert space,
and redundancy, i.e. the disproportionate amount of irrelevant
information encoded in the exact time-dependent wave function.
As a way to bypass this problem, heuristic adiabaticity
conditions \cite{Messiah} inspired by Landau and Zener's work on a two-level model
\cite{landau1932theorie,zener1932non}
have been in use for several decades. The popularity of these conditions is due to
their simplicity, intuitive appeal and reliance on
a small set of physical characteristics of a system.
Unfortunately, these heuristic conditions
were shown to fail even in elementary models
\cite{marzlin2004inconsistency,
tong2005quantitative}. Despite subsequent
progress in mathematical theory of adiabatic processes
\cite{jansen2006bounds,lidar2009adiabatic,bachmann2016adiabatic} the relationship
between the adiabaticity conditions and simple physical
characteristics of a system remains largely unexplored.
Here we show how this gap can be bridged for a broad natural class of
physical systems, that is many-body systems where a small move in
the parameter space induces the orthogonality catastrophe.
In this class, the adiabaticity loss rate has simple expression
in terms of the scaling properties of the parameter-dependent
ground state without a reference to the the excitation spectrum.
This greatly simplifies theoretical investigation of the adiabaticity conditions
by reducing a complex time-dependent problem in a large Hilbert space to the
analysis of the ground state only.

We begin our analysis by noticing that new insight into the
problem of adiabaticity can be obtained by enriching
the general linear-algebraic construction of quantum mechanics
with some additional structure.
Such a structure appears naturally in many body
systems, where the system size plays a role of
an additional control parameter. Known examples
of solvable driven many-body systems
\cite{polkovnikov2008breakdown,altland2008many,
altland2009nonadiabaticity,
shytov2004landau} point to the importance of
this parameter for adiabaticity, although
its general role is not yet understood and
in some cases is a matter of debate
\cite{Lychkovskiy2014,schecter2014comment,Gamayun2014reply}.
To make further progress, we focus on a particular class
of many-body systems where a small move in the parameter
space induces a generalised orthogonality catastrophe.
We define the latter as a phenomenon  by which the overlap $\mathcal C(\lambda) \equiv
\vert \langle  \Phi_{\lambda}
 \vert \Phi_{0}\rangle \vert^2$ has the following asymptotic behaviour  in the
limit of a large particle number $N$
 \begin{equation}
  \ln \mathcal C(\lambda) = - C_N \lambda^2 + r(N, \lambda).
  \label{CNdef}
 \end{equation}
Here $C_N\to \infty$ in the $N\to \infty$ limit, and $r$ is the residual term
satisfying $\lim\limits_{N\rightarrow\infty}r(N, C_N^{-1/2}) = 0$.


We note that the class of many-body systems experiencing the orthogonality
catastrophe in the form \eqref{CNdef} is extremely wide. The theory of the orthogonality
catastrophe is well developed providing efficient tools for the calculation of
$C_N$ such as the linked cluster expansion, effective field theory
methods, variational and Monte Carlo techniques
\cite{Mahan2000,gull2011continuous}. These approaches have been underpinned by
rigorous mathematical results for independent fermion systems
\cite{gebert2014anderson,gebert2014exponent}. It is worth noting that
field-theoretical approaches to the calculation of $C_N$ exploit the method of
adiabatic evolution along the lines of the Gell-Mann and Low theorem.
This requires extra care with taking the thermodynamic
limit  \cite{rivier1971exact,hamann1971orthogonality,kaga1978orthogonality,janis1997complete}.
We emphasise that adiabatic
evolution in such context is a formal device unrelated to any actual physical process.
We further notice that in certain cases $C_N$ can be linked to
a direct experimental measurement,
e.g. to the structure of the X-ray edge singularity \cite{Mahan2000}. Here we take
equation \eqref{CNdef} for granted and proceed to its implications for adiabaticity.

\begin{figure}
 \includegraphics[width = 0.43 \textwidth]{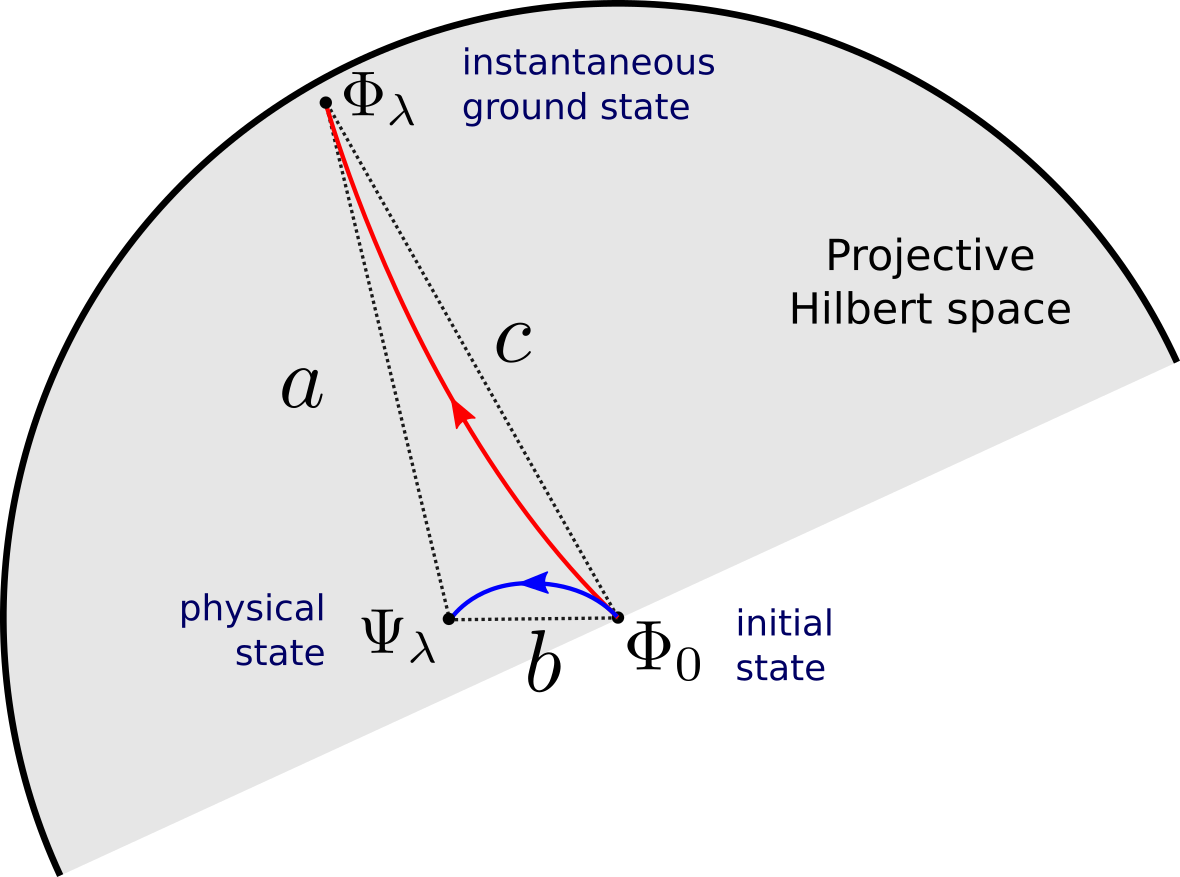}
 \caption{Triangle inequality resulting in the estimate Eq.~\eqref{inequality}. States are shown as
 points in the projective Hilbert space. The red trajectory shows the evolution of the
 instantaneous ground state, Eq.~\eqref{eigenproblem}, while the blue trajectory corresponds to the physical
 evolution given in  Eq.~\eqref{Schrodinger equation}. The length of the side $b$ is bounded by
 the quantum speed limit, while the length of the side $c$ approaches the
 maximally possible distance of $1$ in the large $N$ limit. \label{fig}}
\end{figure}

Our main result in its simplest and most useful form can be stated as follows.
Consider a quantum system with a time-dependent Hamiltonian $\H_\lambda$,
which possesses the following properties
\begin{itemize}
\item[(i)] The system exhibits a generalised orthogonality catastrophe of the form
\eqref{CNdef} with $C_N\to \infty$ in the $N\to \infty$ limit.
\item[(ii)] The uncertainty $\delta V_N \equiv \sqrt{\langle \hat V^2 \rangle_0
- \langle   \hat V \rangle_0^2}$ of the driving potential
$$
\hat V\equiv \left.\frac{\partial \hat H }{\partial \lambda}\right\vert_{\lambda=0}
$$
in the initial state $\Phi_0$  satisfies
\begin{equation}
  \frac{\delta V_N}{C_N} \to 0, \qquad N\to \infty
  \label{cond2}
\end{equation}
\item[(iii)] The fidelity, \eqref{Fdef}, is a monotonically decreasing function of time \footnote{up to Poincare revivals. Their time scale increases so rapidly with increasing system size that they can be safely ignored in
in both practical and asymptotic sense.},
therefore one can define the adiabatic mean free path $\mfp$
as the solution to $\mathcal F(\lambda_*)=1/e$.

\end{itemize}
Then we find that for a driving rate
$\Gamma$ independent of the system size the adiabatic mean free path tends to zero in the $N\to \infty$ limit
with the leading asymptote given by
\begin{equation}
 \mfp = C_N^{-1/2}.
 \label{mainresult}
\end{equation}
It follows that for any fixed
driving rate $\Gamma$ and any fixed displacement $\lambda$
adiabaticity fails if $N$ is large enough to ensure
$\lambda>\lambda_*.$ To avoid the adiabaticity breakdown one has to allow the driving rate
to scale down with increasing system size, $\Gamma=\Gamma_N$, where
\begin{equation}
 \Gamma_N \leq \frac{\delta V_N}{2 C_N }
 \label{necessary}
\end{equation}
in the large $N$ limit.


Next we sketch our derivation of the asymptotic formula for the mean free path, Eq.\eqref{mainresult},
and the necessary condition for a given process to be adiabatic in a large many-body system, Eq.\eqref{necessary}.
They both follow from a rigorous inequality
\begin{align}
\label{inequality}
|{\cal F}(\lambda)-{\cal C}(\lambda)| & \leq
{\cal R}(\lambda)~~~~{\rm with}\nonumber \\
{\cal R}(\lambda') & \equiv
 \int\limits_0^{\lambda'}
 \frac {d\lambda}{|\dot{\lambda}|}
\sqrt{
\la\Psi_0|\H_{\lambda}^2|\Psi_0\ra- \la\Psi_0|\H_{\lambda}|\Psi_0\ra^2
}.
\end{align}
Here $\lambda$ can be an arbitrary smooth function of time and $\dot{\lambda}$
is its derivative. In the large $N$ limit and for $\dot{\lambda}=\Gamma$ the
leading asymptote for ${\cal R}(\lambda)$ is $\lambda^2 \delta V_N /(2\Gamma)$.
Thus the inequality \eqref{inequality} implies that the fidelity $\mathcal F$
and the orthogonality overlap
function $\mathcal C$  stay close to each other for a certain path length
determined
by the ground state uncertainty of the driving potential $\delta V_N$ and the
driving rate $\Gamma$.
 When the system has travelled
the distance $\lambda_*$ given in Eq.~\eqref{mainresult},
then $\mathcal C$ departs significantly from its inital value $\mathcal C=1,$
according to Eq. \eqref{CNdef}.
If at the same time the right hand side of \eqref{inequality}
is still small, which is ensured by Eq.~\eqref{cond2} in the case of a fixed
$\Gamma$,
the fidelity $\mathcal F$ will follow $\mathcal C$ and the adiabaticity
will be lost. This adiabaticity breakdown can be avoided if one allows the
driving rate to scale with the system size.
In this case one imposes the condition \eqref{necessary} to ensure that  the
r.h.s. of Eq. \eqref{inequality} is greater than one and thus
$\mathcal F$ and $\mathcal C$ are unrelated.

The proof of the inequality \eqref{inequality} is given in the supplement
\cite{supplementary}. Here we outline the main idea behind this proof.
Firstly, we recall that the space of quantum states can be endowed by a
natural sense of distance, the Bures angle distance \cite{bengtsson2007geometry}
\begin{equation}\label{Bures angle}
 D(\Phi, \Psi) = \frac2\pi \arccos|\langle \Phi \vert  \Psi \rangle|,
\end{equation}
where $\Phi$ and $\Psi$ are any two states represented by
normalised wave functions in the system's Hilbert space.
As the parameter $\lambda$ changes the physical state
$\Psi_\lambda$ and the instantaneous ground state $\Phi_\lambda$
each describe a continuous trajectory in this metric space as illustrated
in Fig.~\ref{fig}. At any given $\lambda$ the states $\Phi_0,$
$\Psi_\lambda$ and $\Phi_\lambda$ form a triangle with sides
$a,$ $b$ and $c.$ The sides $a$ and $c$ characterise the
fidelity and the orthogonality overlap respectively. In order
to estimate the side $b$ we employ the quantum speed limit
\cite{pfeifer1993fast,pfeifer1995generalized}, which provides an upper bound on the length of
$b$ in terms of the quantum uncertainty of the driving
potential. The inequality \eqref{inequality} then follows
from the triangle inequality $|a-c|\leq b$.

Next, we discuss, without going too deeply into the mathematical detail,
the scaling properties of $C_N$ and $\delta V_N$ and explain why
applicability conditions (i) and (ii) hold in a broad class of many-body
systems. For simplicity, we  limit ourselves to the case of a standard thermodynamic
limit taken at a fixed particle density.
We recall that typical physical observables in a many-body system are
generated by quasi-local operators having a finite-range support in the
configuration space.  We denote one such operator $\hat v(x)$ where $x$ is a
point in a $D$-dimensional space and take
\begin{equation}
 \hat V = \int_{\text{vol}} d^Dx f(x) \hat v(x),
\end{equation}
where the integral is taken over the volume of the system and $f(x)$
is a support function, which satisfies
\begin{equation}
  \int_{\text{vol}} d^Dx f(x) \sim N^{d/D}, \qquad N\to \infty.
\end{equation}
For example, if $f(x)$ constraints driving to
the boundary of the sample we have $d=D-1$, for driviing localized
near a given point of space we have $d=0,$
while for driving homogeneously distributed in the bulk we have $d=D.$
It is straightforward to see that in {\it all} systems with rapidly
decaying local correlation functions, for example, in systems with
a spectral gap,  $\delta V_N \sim N^{d/(2D)}$ while $C_N\sim N^{d/D},$
which immediately ensures conditions (i) and (ii) for $d>0$.
For localized driving, $d=0$, conditions (i) and (ii) are violated unless
the spectrum of the system is gapless. For example, in a metal
$C_N \sim \log N$ \cite{anderson1967infrared,gebert2014anderson}
(other scaling laws may apply in dirty metals
\cite{gefen2002anderson,Kettemann2016} or near quantum critical points
\cite{polkovnikov2011universal}) and $V_N \sim 1.$

To illustrate our general findings, we consider the Rice-Mele model, describing a system of
fermions on a half-filled one-dimensional bipartite lattice with
the Hamiltonian
\begin{multline}
 H_{\rm RM}=
 \sum_{j=1}^N \left[-(J + U) a_j^\dagger b_j
 - (J - U) a_j^\dagger b_{j+1} + \mathrm {h.c.} \right]
 \\ + \sum_{j=1}^N \Delta (a^\dagger_j a_j - b^\dagger_j b_j ) .
 \label{R-M}
\end{multline}
Here $a_j$ and $b_j$ are the fermion annihilation operators on the
$a$ and $b$ sublattices, and $j$ labels the lattice cites.
The Rice-Mele Hamiltonioan is an archetypal model of the adiabatic
Thouless pump, that is a system where exactly one particle is
transferred from one end to another if a topologically
non-trivial cycle is performed in the Hamiltonian's parameter
space \cite{thouless1983quantization}.
In the present case such a cycle would be any loop in the $(U, \Delta)$
plane enclosing the origin. The reasoning of \cite{thouless1983quantization}
guarantees quantization of the pumped charge provided
the evolution is adiabatic, however the adiabatic conditions are
not elaborated upon in \cite{thouless1983quantization}.




\begin{figure}
\includegraphics[width=0.45 \textwidth]{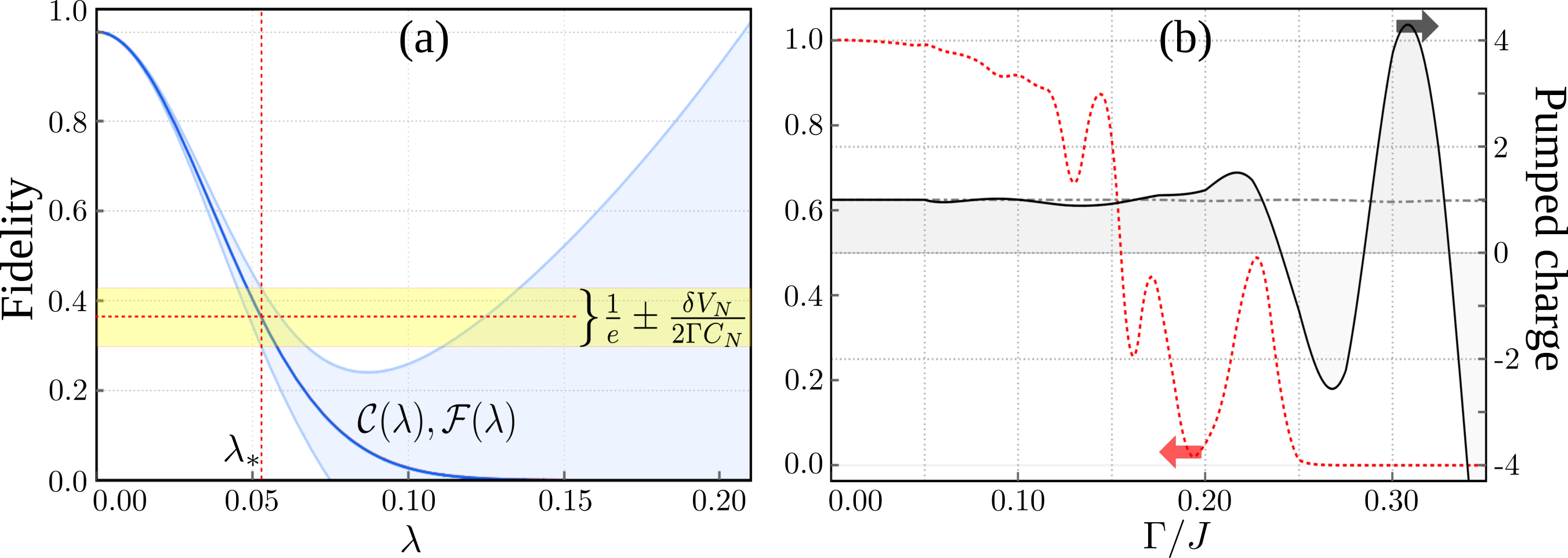}
\caption{Adiabaticity, orthogonality catastrophe and transport in a Thouless pump described by the Hamiltonian \eqref{R-M} with $N=1000$ particles. (a) Illustration of the inequality \eqref{inequality}. Solid blue -- the orthogonality overlap $\mathcal C(\lambda)$. The shaded region is the one, which has to contain the
$\mathcal F(\lambda)$ curve due to the inequality \eqref{inequality}. This bound is tight enough to guarantee the adiabaticity breakdown for the chosen set of parameters, $J=0.4 E_R,$ $ U = 0.4 E_R$, $\Delta = \lambda E_R$ with
$\lambda = \Gamma t$ and $\Gamma= 0.7 E_R$. For $E_R= 6.4$ ms$^{-1}$ these parameters coincide with those of the effective Hamiltonian describing the optical lattice in the experiment Ref.~\cite{Nakajima2016topological}
at the $\Delta = 0$ point of the pumping cycle. Remarkably, true adiabatic fidelity $\mathcal F (\lambda)$ calculated numerically follows $\mathcal C(\lambda)$ even closer than what can be expected from the bound  \eqref{inequality}, so that the two curves look indistinguishable in the figure. The latter fact indicates that the sides $b$ and $c$ of the triangle depicted in Fig.~\ref{fig} are in fact nearly orthogonal in the present case.
(b) The  charge transferred during the cycle (dash-dotted gray) as compared to  the total transferred charge collected until the current vanishes (solid black) for different values of  the driving rate, $\Gamma$. The cycle is given by $\Delta=(J/2)\sin\lambda$, $U=(J/2)\cos\lambda$ with
$\lambda=\Gamma t$. The pump performs a single cycle and then stops. Initially the system is in equilibrium.  The adiabatic fidelity
$\mathcal F$ at the end of the cycle is shown by the dashed red line.       One can see that the charge transferred during the cycle is quantized even when the
many-body adiabaticity has gone completely ($\mathcal F\simeq0$), while the quantization of the total transferred charge 
disappears as soon as the many-body adiabaticity is broken.
\label{Thouless}
}
\end{figure}

\footnotetext{blah blah blah}

The Thouless pump protocol in a Rice-Mele system was
recently implemented in a parabolically confined ultracold atomic system
\cite{Nakajima2016topological} (see also \cite{Lohse2015thouless}
for pumping in a Bose-Mott insulator). The particle transport was measured
by the direct observation of the centre-of-mass displacement of the atomic cloud
using {\it in situ} absorption imaging.
The authors of Ref.~\cite{Nakajima2016topological} emphasise the
importance of adiabaticity for the observation of the Thouless
quantisation. In order to ensure slow enough driving, they use a
heuristic condition $\Gamma < \Gamma_{\rm LZ}$.
Here $\Gamma_{\rm LZ}$ is obtained in the Landau-Zener spirit from the
condition $2\pi(D_{\rm min}/2)^2/(\dot{D}_{\rm max})=1$,
where  $D_{\rm min}$ is the smallest value of the time-dependent band gap,
$D(t)$, and $\dot{D}_{\rm max}$ is the maximal derivative of $D(t)$  during the
cycle.
We note that this condition is insensitive to the system size,
in particular it does not predict any problems with
adiabaticity in the thermodynamic limit.
In contrast, our exact result \eqref{mainresult}, together with
the scaling laws $C_N\sim N$ and $\delta V_N\sim \sqrt N$
(see \cite{supplementary}) imply
that for any given $\Gamma<\Gamma_{\rm LZ}$
adiabaticity fails to survive even a single cycle of pumping
when the number of particles is too large \footnote{This scaling exemplifies a remarkable fact: In a many-body system a finite
energy gap is not sufficient to protect adiabaticity in the thermodynamic limit.
This fact can be easily inferred from our general scaling analysis for $d \geq
1$.}. To illustrate the effect of the system size on adiabaticity we
numerically simulate the evolution of the fidelity $\mathcal F$
in the Rice-Mele model \eqref{R-M}for various system sizes,
with parameters $J$, $U$, $\Delta$ taken from the experiment \cite{Nakajima2016topological}.
While for $N=10$, which
is close to the experimental value,
the bound \eqref{inequality} is too weak to relate the
fidelity to the orthogonality catastrophe \cite{supplementary},
for $N=1000$ the bound is strong enough to ensure an
$e$-fold decay of the fidelity over a mean-free path,
which turns out to be only a small fraction of the complete cycle, see Fig. \ref{Thouless} (a).

It is instructive to investigate
what happens to the Thouless pumping
in the regime where
the particle number is large enough to ensure the
collapse of adiabaticity within one cycle, but the driving is slow compared to the bandgap.
First, we consider this
question qualitatively assuming, for simplicity, a two-terminal
geometry where the ends of the Rice-Mele lattice are attached to two
infinite particle reservoirs. In such a geometry the charge is pumped
between the two reservoirs. We recall that the adiabatic mean free path
$\lambda_* \sim 1/\sqrt N $ is the typical
distance the Thouless pump travels in the parameter space
before an elementary excitation is created in the bulk. For $\lambda_*$
much shorter than the length of the loop that the system describes
in the parameter space, a  large number of elementary
excitations is born in one  cycle. These excitations form a
dilute gas of mobile quasi-particles, which travel in both
directions, left and right. If the pump is initiated in the equilibrium
state and performs one cycle, then during the period $T$ of the cycle the
number of such excitations reaching the left/right end of the system
will be $\delta N = \rho v T,$ where $\rho\sim 1/(\lambda_* N)$
is the number of elementary excitations created during the cycle per lattice site and $v$ is the typical group
velocity. Clearly, $\delta N \ll 1$ in the large $N$ limit, therefore
the charge pumped in one cycle will be close to the
quantized value despite the violation of adiabaticity conditions.
The quantization also survives in the steady regime of operation of the pump, when it performs one cycle after another. This has been demonstrated in Ref. \cite{Privitera_2018} by analyzing the Floquet eigenstates of the pump.

However, the quantization breaks down simultaneously with adiabaticity in a yet different setting, when the pump performs a single cycle and stops, and one counts  {\it all } the transferred charge, until the current vanishes (which happens long after the cycle ends). In this case all $\sim \sqrt{N}$ quasiparticles will eventually reach the reservoirs thus destroying the quantization.  This conclusion is supported by a microscopic calculation (given in the Supplement) as illustrated in Fig. \ref{Thouless} (b).

To conclude, we have established a simple quantitative relationship between the
orthogonality catastrophe and the adiabaticity breakdown in a driven many-body
system. We have illustrated the utility of this finding by determining conditions for quantization of transport in a Thouless pump.

\begin{acknowledgments}
{\it Acknowledgements.} The authors are grateful to P. Ostrovsky, S. Kettemann,
I. Lerner, G. Shlyapnikov, Y. Gefen and M. Troyer for fruitful discussions and
useful comments, and to S. Nakajima for clarifying
the experimental conditions of Ref.  \cite{Nakajima2016topological}. OL
acknowledges the support from the Russian Foundation for Basic Research
under Grant No. 16-32-00669. The work of OG was partially supported by Project 1/30-2015  ``Dynamics and topological structures in Bose-Einstein condensates of ultracold gases''  of the KNU  Branch Target Training at the NAS of Ukraine.
\end{acknowledgments}

\setcounter{equation}{0}

\renewcommand{\theequation}{S\arabic{equation}}

\begin{widetext}

\appendix

\begin{center}
{\Large Supplement}
\end{center}

\section{{\large\bf 1.~} Quantum speed limit and relation between orthogonality catastrophe and adiabaticity}

In contrast to the main text, in the present section we use time, not $\lambda$, to parameterize the instantaneous ground state of the system, $\Phi_t$, and the evolving state of the system, $\Psi_t$. The former is a solution of the Schrodinger's stationary equation
\be\label{eigenproblem supp}
\H_{\lambda(t)}\, \Phi_t = E_{\lambda(t)} \, \Phi_t,
\ee
while the latter satisfies the Schrodinger's equation
\be\label{Schrodinger equation supp}
i\,\frac\partial{\partial t} \Psi_t = \H_{\lambda(t)}\,
\Psi_t
\ee
with the initial condition $\Psi_0=\Phi_0$. Here $\lambda(t)$ can be an arbitrary smooth function of time. We will also slightly abuse the notations and write ${\cal R}(t)\equiv{\cal R}\left(\lambda(t)\right)$.

\subsection{\label{sec inequality} {\large \bf 1.1.~} Relation between orthogonality catastrophe and adiabaticity}

Here we prove the inequality (8) of the main text which relates the orthogonality overlap $\cal C(\lambda)$ with the adiabatic fidelity $\cal F(\lambda)$. We rewrite it as follows:
\begin{equation}
\label{inequality general}
|{\cal F}\left(\lambda(t)\right)-{\cal C}\left(\lambda(t)\right)| \leq {\cal R}(t)\equiv  \int_0^t \sqrt{
\la\Psi_0|\H_{\lambda(t')}^2|\Psi_0\ra- \la\Psi_0|\H_{\lambda(t')}|\Psi_0\ra^2 }\, dt'.
\end{equation}
Here the integration is performed over the path in the parameter space parameterised by time, and $t$ corresponds to the end point $\lambda$ of this path.

In order to prove the bound \eqref{inequality general} we employ the quantum speed limit (QSL) in the following form:
\be\label{QSL}
 D(\Psi_0,\Psi_t)~\leq~ \frac2\pi \, {\cal R}(t),
\ee
where $D$ defined by Eq. (9) of the main text
is a distance on the Hilbert space known as Bures angle, quantum angle or Fubini-Study metric.
The QSL \eqref{QSL} is a direct consequence of a more general result by Pfeifer \cite{pfeifer1993fast,pfeifer1995generalized}. A detailed derivation of eq. \eqref{QSL} can be found in the next subsection.

Combining the QSL \eqref{QSL} with the triangle inequality
\be
|D(\Phi_t,\Psi_0)-D(\Phi_t,\Psi_t)|~\leq~D(\Psi_0,\Psi_t)
\ee
and taking into account that $\Psi_0=\Phi_0$, one gets
\be\label{pre-ours inequality}
|D(\Phi_t,\Phi_0)-D(\Phi_t,\Psi_t)|~\leq~ \frac2\pi {\cal R}(t).
\ee

Finally, one obtains the inequality \eqref{inequality general} from the inequality \eqref{pre-ours inequality} by observing that
\be\label{algebraic}
|x^2-y^2| ~\leq ~ |\arccos x-\arccos y|~~~{\mathrm {for ~all}}~~~|x|\leq1,~~|y|\leq1.
\ee

\vspace{1.5 em}
One may wonder what is the reason for using the Bures angle distance instead of e.g. a more conventional trace distance, $D_{\rm tr}(\Psi,\Phi)\equiv\sqrt{1-\left|\la \Phi|\Psi\ra\right|^2}$ . It is easy to see that the trace distance is bounded by the Bures angle, $D_{\rm tr}(\Psi,\Phi)<(\pi/2)D(\Psi,\Phi)$, and thus eq. \eqref{QSL} entails the following (weaker) version of the QSL,
\be\label{QSL 2}
D_{tr}(\Psi_0,\Psi_t)~\leq~{\cal R}(t).
\ee
However, if we try to move forward with this QSL instead of eq. \eqref{QSL}, we get an extra factor 2 in the r.h.s. of the bound \eqref{inequality general}. Let us show this. Using  \eqref{QSL 2} and triangle inequality for the trace distance, one obtains an analog of \eqref{pre-ours inequality}:
\be\label{pre-ours inequality 2}
|D_{tr}(\Phi_t,\Phi_0)-D_{tr}(\Phi_t,\Psi_t)|~\leq~{\cal R}(t).
\ee
Now one has to relate the l.h.s. of this inequality with the l.h.s. of inequality \eqref{inequality general}. This amounts to relating $|\sqrt{1-x^2}-\sqrt{1-y^2}|$ with $|x^2-y^2|$, and at this point extra 2 emerges. This is because one can only guarantee that
\be
|x^2-y^2|\leq 2|\sqrt{1-x^2}-\sqrt{1-y^2}|,
\ee
compare to eq. \eqref{algebraic}.

\subsection{{\large \bf 1.2.~} Quantum speed limit}

Here we derive the QSL limit \eqref{QSL} from a result by
Pfeifer  \cite{pfeifer1993fast,pfeifer1995generalized} which reads

\begin{align}\label{inequality Pfeifer}
\sin_* \left( \arcsin\left|\la F|\Psi_0\ra\right| - {\cal R}(t) \right)~ \leq ~|\la F|\Psi_t\ra|
~\leq~ \sin_* \left( \arcsin\left|\la F|\Psi_0\ra\right| + {\cal R}(t) \right).
\end{align}
Here $F$ is an arbitrary auxiliary state and
\be
\sin_* x \equiv
\left\{
\begin{array}{ll}
0, & x<0,\\
\sin x, & 0\leq x \leq \pi/2,\\
1, & x>\pi/2.
\end{array}
\right.
\ee
 Noting that $\arcsin x+\arccos x=\pi/2$ and $\sin_*(\pi/2- x)= \cos_* x$ with
\be
\cos_* x \equiv
\left\{
\begin{array}{ll}
1, & x<0,\\
\cos x, & 0\leq x \leq \pi/2,\\
0, & x>\pi/2,
\end{array}
\right.
\ee
one  can rewrite \eqref{inequality Pfeifer} as
\begin{align}\label{inequality Pfeifer 2}
\cos_* \left( \arccos\left|\la F|\Psi_0\ra\right| + {\cal R}(t) \right) ~\leq~ |\la F|\Psi_t\ra|
~\leq~ \cos_* \left( \arccos\left|\la F|\Psi_0\ra\right| - {\cal R}(t) \right).
\end{align}
Taking into account that
\be
\arccos(\cos_*x)=
\left\{
\begin{array}{ll}
0, & x<0,\\
x, & 0\leq x \leq \pi/2,\\
\pi/2, & x>\pi/2,
\end{array}
\right.
\ee
one rewrites eq. \eqref{inequality Pfeifer 2} in terms of the Bures angle:
\begin{align}\label{inequality Pfeifer 3}
\max\{D(F,\Psi_0)-\frac2\pi\,{\cal R}(t),0\}
~\leq~ D(F,\Psi_t)~\leq~
\min\{D(F,\Psi_0)+\frac2\pi\,{\cal R}(t),1\}.
\end{align}
Employing  obvious relations $\min\{x,y\}\leq x$ and $\max\{x,y\}\geq x$ one reduces \eqref{inequality Pfeifer 3} to a more compact, though slightly more rough inequality,
\be\label{inequality Pfeifer 4}
\left|D(F,\Psi_t)-D(F,\Psi_0)\right|~\leq~\frac2\pi\,{\cal R}(t).
\ee
Choosing $F=\Psi_0$ one obtains the QSL \eqref{QSL}.

It should be noted that another choice, $F=\Phi_t$, directly reduces the inequality  \eqref{inequality Pfeifer 4} (along with the condition $\Psi_0=\Phi_0$) to the inequality \eqref{pre-ours inequality}.  Such a direct route which apparently dispenses with the triangle inequality is possible because Pfeifer's rather sophisticated result has, in fact, a more broad scope than elementary versions of the quantum speed limit and contains the  triangle inequality built in.

\section{{\large\bf 2.~} Rice-Mele model}
\subsection{{\large \bf 2.1.~} Eigenstates and eigenenergies}
The transformation
\begin{align}
a_j=\frac1{\sqrt N}\sum_k e^{ikj}a_k,~~b_j=\frac1{\sqrt N}\sum_k e^{ikj}b_k,~~k=\frac{2\pi}{N}l, ~~l=-\frac{N}2+1,-\frac{N}2+2,...,\frac{N}2,
\end{align}
where $N$ is assumed to be even,
allows one to  represent the Rice-Mele Hamiltonian, eq. (12) in the main text,  as a sum of $N$ commuting terms,
\be\label{R-M 2}
\H_{\rm RM}=\sum_k
(a^\dagger_k~b^\dagger_k)
\left(
\begin{array}{c c}
\Delta & -(J+U)-(J-U)e^{ik}\\
-(J+U)-(J-U)e^{-ik} & -\Delta
\end{array}
\right)
\left(
\begin{array}{c }
a_k\\
b_k
\end{array}
\right).
\ee
Observe that $n_k \equiv a^\dagger_k a_k+b^\dagger_k b_k$ is conserved for each $k$. We assume  half-filling, i.e. that the total number of particles equals $N$. In this case the ground state of the Hamiltonian for any values of $(J,U,\Delta)$ is an eigenstate of $n_k$ with the eigenvalue equal to 1, and this is maintained throughout the evolution. Restricting the Hamiltonian \eqref{R-M 2} to the corresponding subspace one obtains an effective Hamiltonian of $N$ noninteracting spins,
\be\label{R-M 3}
\H_{\rm RM}=\sum_k {\pmb p_k}\cdot\ssigma_k,~~~~~~~~~{\pmb p_k}=
\left(
\begin{array}{c }
-(J+U)-(J-U)\cos k\\
(J-U)\sin k\\
\Delta
\end{array}
\right),
\ee
where $\ssigma$ is a vector consisting of three Pauli matrices. Each two-level Hamiltonian $\H_k \equiv {\pmb p_k}\cdot\ssigma_k$ has two eigenstates $|\chi_k^\pm\ra$ and eigenenergies $\varepsilon_k^\pm$,
\be\label{eigensystem}
\rho_k^\pm\equiv|\chi_k^\pm\ra\la\chi_k^\pm|=\frac12(1\pm\frac1{|\pmb p_k|}\,{\pmb p_k}\cdot\ssigma_k), ~~~~~~~~~~\varepsilon_k^\pm=\pm\sqrt{2(J^2+U^2)+\Delta^2 + 2(J^2-U^2)\cos k}.
\ee
The ground state of the whole system is the product of $N$ single-spin eigenstates, $|\Phi\ra=\prod\limits_k \otimes |\chi_k^-\ra$ while the ground state energy  is the sum of corresponding eigenenergies,  $E=\sum\limits_k\varepsilon_k^-$.

\subsection{{\large \bf 2.2.~} Orthogonality catastrophe}

Here we consider the orthogonality catastrophe induced by changing the parameters of the Hamiltonian $(J,U,\Delta)$ along some trajectory  parameterized by $\lambda$. This is to say that $(J,U,\Delta)$ and thus vectors $\pmb p_k$ are functions of $\lambda$. In contrast to the main text, we do not employ the convention that $\lambda=0$ at $t=0$.

The orthogonality overlap for a single spin reads
\be
\mathfrak{c}_k (\lambda',\lambda)\equiv |\la\chi_k^-(\lambda')|\chi_k^-(\lambda)\ra|^2=\tr(\rho_k^-(\lambda)\rho_k^-(\lambda')) = \frac12(1+\frac{\p_k(\lambda) \cdot \p_k(\lambda')}{|\pmb p_k(\lambda)|\, |\pmb p_k(\lambda')|}).
\ee
The  orthogonality overlap for the whole many-body system is given by
\be\label{orthogonality overlap}
\mathcal{C}(\lambda',\lambda;N)\equiv|\la\Phi_{\lambda'}|\Phi_\lambda\ra|^2 = \exp\left(-\sum_k \log\frac1{\mathfrak{c}_k}\right).
\ee
%
%
Further,
\be\label{CN R-M general}
C_N=\sum_k \, c_k~~~{\rm with} ~~~
c_k\equiv -\frac12
\left.
\left(
\frac{\partial^2}{\partial \lambda^2} \log \mathfrak{c}_k(\lambda',\lambda)
\right)
\right|_{\lambda'=\lambda}=\frac14\left( \frac{\partial_\lambda \p_k\cdot\partial_\lambda \p_k}{|\p_k|^2}-\frac{(\p_k\cdot\partial_\lambda \p_k)^2}{|\p_k|^4}  \right).
\ee

This equation along with eq. \eqref{R-M 3} enables one to calculate $C_N$ for any point of any trajectory in the parameter space of the Rice-Mele model. For example, for $J,U={\rm const}$, $\Delta=\lambda E_R$ one obtains
\be
c_k=\frac18 \frac1{ J^2 + U^2 + (J^2 - U^2) \cos k}
\ee
and
\begin{equation}\label{CN R-M spec}
C_N=\frac{N E_R^2}{16 J U}.
\end{equation}

\subsection{{\large \bf 2.3.~} Quantum uncertainty of the driving potential}

To deal with general trajectories we define the driving term as
\be
\V\equiv \partial_\lambda \H_\lambda,
\ee
which is consistent with the definition adopted in the main text. For the Rice-Mele model
\be
\V=\sum_k \partial_\lambda \p_k\cdot\ssigma_k.
\ee
Since the ground state is of the product form, the quantum uncertainty of $\V$ is expressed through individual uncertainties of states of single spins:
\be
\delta V^2=
\sum_k
\left(
 \tr\left( \rho_k^-\,(\partial_\lambda \p_k\cdot\ssigma_k)^2  \right)
 - \left(\tr\left( \rho_k^-\,\partial_\lambda \p_k\cdot\ssigma_k \right)\right)^2
\right)
=
\sum_k
\left(
|\partial_\lambda \p_k|^2
 -\frac{(\p_k\cdot\partial_\lambda \p_k)^2}{|\p_k|^2}
\right).
\ee
This equation along with eq. \eqref{R-M 3} allows one to calculate $\delta V_N$ for any point of any trajectory in the parameter space of the Rice-Mele model. For example, for $J,U={\rm const}$, $\Delta=\lambda E_R$ one gets
\begin{equation}\label{delta VN R-M spec}
\delta V_N^{\rm RM}=\sqrt N E_R.
\end{equation}

With the knowledge of $C_N$ and $\delta V_N$ one can make practical use of the inequality (8) of the main text, or, alternatively, inequality  \eqref{inequality general}. For a fixed $\Gamma$ the r.h.s. of this inequality inevitably diminishes with growing $N$, leading to $\mathcal F (\lambda)\simeq \mathcal C (\lambda)$, as illustrated in Fig. 2 (a) of the main text.  The opposite situation when the r.h.s. of \eqref{inequality general} is large and thus the inequality \eqref{inequality general} is inconclusive is illustrated in Fig. \ref{fig1}.

\begin{figure}[t]
\includegraphics[width=0.45 \textwidth]{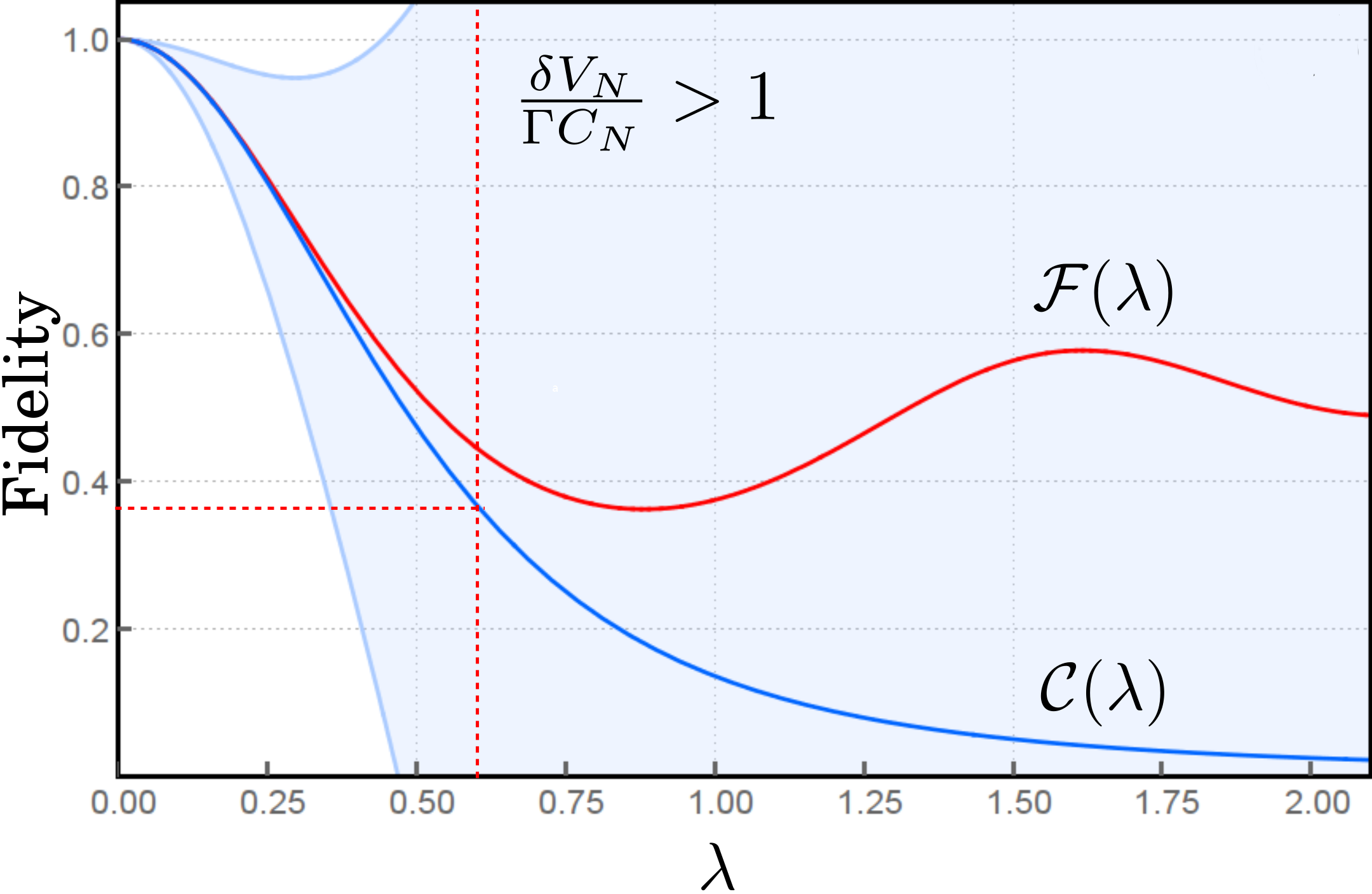}
\caption{Evolution of the ground state fidelity in the parameter space
of the Hamiltonian \eqref{R-M 2} for  $N=10$ particles.
The fidelity $\mathcal F$ is shown as a solid red curve,
while the overlap function $\mathcal C$ is shown as a solid blue curve.
The shaded region is the one, which has to contain the
$\mathcal F(\lambda)$ curve due to the inequality \eqref{inequality general}.
For $N=10$ the inequality \eqref{inequality general} does not impose a meaningful upper bound on the fidelity, and therefore
has nothing to say about the relationship between the fidelity and the orthogonality catastrophe.
The parameters used read
$J=0.4 E_R,$ $ U = 0.4 E_R$ and $\Delta = \lambda E_R$ with
$\lambda = \Gamma t$ and $\Gamma= 0.7 E_R.$
For the recoil energy $E_R= 6.4$ ms$^{-1}$ these coincide with the
parameters of the effective Hamiltonian describing the optical
lattice in the experiment Ref.~\cite{Nakajima2016topological}
at the $\Delta = 0$ point of the pumping cycle.
\label{fig1}
}
\end{figure}

\subsection{{\large \bf 2.4.~} Current and transferred charge}

The current flowing between the $l$'th  and $(l+1)$'th elementary cell reads
\be
\hat j_l=i(J-U)\sum_{l=1}^N (b_{l+1}^\dagger a_l-a_l^\dagger b_{l+1}).
\ee
Due to translation invariance of the Hamiltonian and the initial state the current is the same for all cells. It is convenient to define an average current,
\be
\hat j\equiv\frac1N \sum_{l=0}^N \hat j_l,
\ee
and then express it in terms of $a_k$, $b_k$:
\be
\hat j=\frac{i}N \sum_k
(a^\dagger_k~b^\dagger_k)
\left(
\begin{array}{c c}
0 & -(J-U)e^{ik}\\
(J-U)e^{-ik} & 0
\end{array}
\right)
\left(
\begin{array}{c }
a_k\\
b_k
\end{array}
\right).
\ee
In terms of spin variables the current can be written as
\be
\hat j=\frac1N\sum_k\hat j_k~~~{\rm with}~~~ \hat j_k= {\pmb j_k}\cdot\ssigma_k,~~~~~~{\pmb j_k}=
\left(
\begin{array}{c }
(J-U)\sin k\\
(J-U)\cos k\\
0
\end{array}
\right),
\ee
The pumped charge is the integral of the quantum average of this current over the elapsed time:
\be\label{Q(t)}
Q(t)=\int_0^t dt'\, \langle \Psi_{t'} |\hat j| \Psi_{t'}\rangle.
\ee
Thanks to eq. \eqref{R-M 3} and factorized initial condition $\Psi_0=\Phi_0=\prod\limits_k \chi^-_k$, eq. \eqref{Q(t)} can be written as
\be
Q(t)=\int_0^t dt' \sum_k \langle \chi_k(t') |\hat j_k| \chi_k(t') \rangle,
\ee
where $\chi_k(t)$ is found from the Schrodinger equation
\be\label{SE factorized}
i\,\partial_t\, \chi_k(t)=({\pmb p_k}(t)\cdot\ssigma_k )\, \chi_k(t).
\ee

\begin{figure}[t]
\includegraphics[width=0.45 \textwidth]{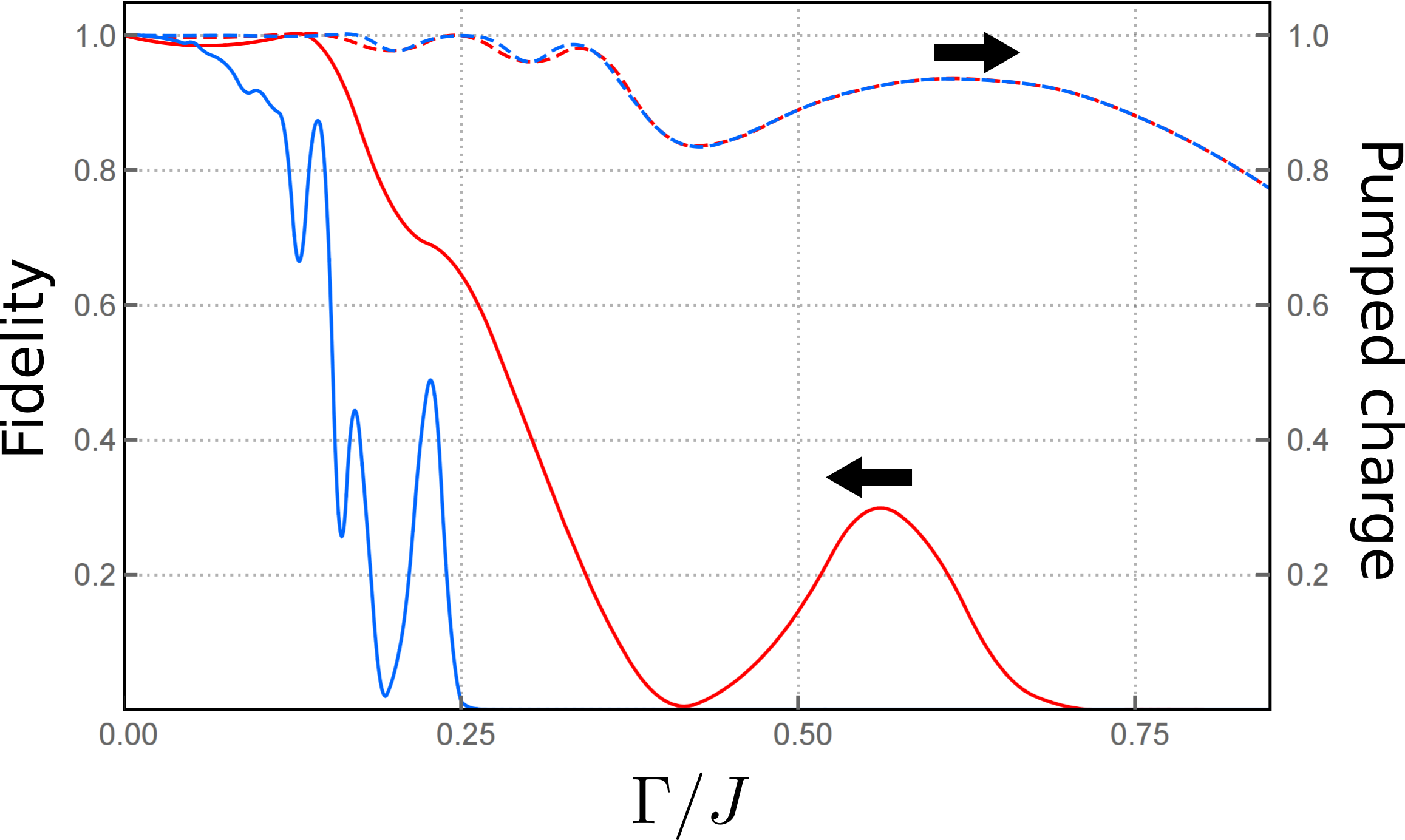}
\caption{\label{fig 2}
Fidelity (solid curves) and  pumped charge (dashed curves) after a single cycle in the
Rice-Mele realisation of the Thouless pump, eq. \eqref{R-M 2}. The trajectory of the cycle
is given by $\Delta=(J/2)\sin\lambda$, $U=(J/2)\cos\lambda$ with
$\lambda=\Gamma t$. The system is initiated in equilibrium. Red and blue curves correspond to  $N=100$ and $N=1000$
fermions in a lattice, respectively. One can see that the
charge transferred in a single cycle hardly depends on the number of particles for $N\gtrsim100$ while the
fidelity decays more rapidly for larger~$N$.
}
\end{figure}

In the context of Thouless pumping we enquire how much charge is transferred per cycle immediately after the  cycle is over, and long after the cycle is over. To answer the former question we calculate $Q(t)$ by solving the Schrodinger equations \eqref{SE factorized} numerically. The result is illustrated in Fig. \ref{fig 2}. The latter question is addressed by counting the number of right- and left- moving excitations produced during the cycle. To this end we define the average population of the excited state with the quasimomentum $k$,
\be
w_k= \langle \chi_k(T) |\frac12(1+{\pmb p_k}(T)\cdot\ssigma_k)| \chi_k(T) \rangle.
\ee
Remind that  $\chi_k(T)$ should be found numerically from the Schrodinger equation \eqref{SE factorized}. Taking into account that ${\pmb p_k}(T)={\pmb p_k}(0)$ this can be rewritten with the use of eq. \eqref{eigensystem} as
\be
w_k= |\langle \chi_k^+ | \chi_k(T) \rangle|^2.
\ee
The sign of the group velocity of the excitations, $\partial(\varepsilon_k^+-\varepsilon_k^-)/\partial k$, coincides with the sign of $k$ for $-\pi<k<\pi$, as is clear from eq. \eqref{eigensystem}. Therefore the charge transferred from left to right per cycle in the steady state, $\Delta Q$, reads
\be
\Delta Q=\sum_{0<k<\pi}w_k-\sum_{-\pi<k<0}w_k.
\ee

\end{widetext}

\bibliography{C:/D/Work/QM/Bibs/1D,C:/D/Work/QM/Bibs/LZ_and_adiabaticity,C:/D/Work/QM/Bibs/orthogonality_catastrophe,C:/D/Work/QM/Bibs/QSL,C:/D/Work/QM/Bibs/QIP,C:/D/Work/QM/Bibs/spin_chains}

\end{document}